\documentclass[amsmath,superscriptaddress,showpacs,aps,twocolumn,prb]{revtex4}
\usepackage{bm}
\usepackage{graphics}
\usepackage{graphicx}
\begin{document}

\title{Electron transport in p-wave superconductor-normal metal junctions. }

\author {A. Keles}

\affiliation{Department of Physics, University of Washington, Seattle, WA 98195}

\author{A. V. Andreev}

\affiliation{Department of Physics, University of Washington, Seattle, WA 98195}

\author {B. Z. Spivak}

\affiliation{Department of Physics, University of Washington, Seattle, WA 98195}

\begin{abstract}

We study low temperature electron transport in p-wave
superconductor-insulator-normal metal junctions. In diffusive metals the p-wave
component of the order parameter decays exponentially at distances larger than
the mean free path $l$. At the superconductor-normal metal boundary, due to
spin-orbit interaction, there is a triplet to singlet conversion of the
superconducting order parameter. The singlet component survives at distances
much larger than $l$ from the boundary. It is this component that controls the
low temperature resistance of the junctions. As a result, the resistance of the
system strongly depends on the angle between the insulating boundary and the
${\bf d}$-vector characterizing the spin structure of the triplet
superconducting order parameter.  We also analyze the spatial dependence of the
electric potential in the presence of the current, and show that the electric
field is suppressed in the insulating boundary as well as in the normal metal at
distances of order of the coherence length away from the boundary.  This is very
different from the case of the normal metal-insulator-normal metal junctions,
where the voltage drop takes place predominantly at the insulator.

\end{abstract}
\date{\today}

\maketitle

\section{Introduction}

Electron transport in  superconducting systems is very different from that in
normal metals.  Roughly speaking, the characteristic size of wave packets which
carry current in metals is of the order of the Fermi wave length $\hbar/p_{F}$,
while their charge is equal to the electron charge $e$.  Here $p_{F}$ is the
Fermi momentum. On the other hand, the quasiparticle wave packets  are coherent
superpositions of electrons and holes.  This results in a characteristic size of
the wave packets which is much larger than $\hbar/p_{F}$.  The charge of the
packets depends on the energy and can be very different from the electron charge
$e$.  This has important consequences in electronic  transport properties of
superconductor-insulator-normal metal junctions.

Transport properties of s-wave superconductor-insulator-normal metal junctions
have been the subject of intensive experimental and theoretical research for
decades, see for example,
Refs.~\onlinecite{Klapwijk,hekking,nazarov_stoof,beenakker1,beenakker,spivak}.
In this case the Cooper pairs can be constructed from the two single particle
wave functions related by a time reversal operation. At low temperatures the
characteristic size of wave  packets which carry current in the metal near the
boundary is of the order of the normal metal coherence length
$L_{T}=\sqrt{D/T}$, which turns out to be much larger than the elastic mean free
path $l$. Here $D$ is the diffusion coefficient, and $T$ is the temperature.
One of the consequences of the large size of the wave packets is that, in the
presence of a current through the junction, the drop of the gauge-invariant
potential $\Phi$ is pushed to distances of order $L_{T}$ away from the boundary,
which is much larger than both the thickness of the insulator and the elastic
mean free path $l$. This is quite different from the case of normal
metal-insulator-normal metal junctions, where most of the potential drop occurs
at the insulator.

In this article we develop a theory of electron transport in p-wave
superconductor-insulator-normal metal junctions.  The best known example of a
p-wave superfluid is superfluid $^{3}He$.   One of the leading candidates for
p-wave pairing in electronic systems is $Sr_{2}RuO_{4}$.  There are numerous
pieces of experimental evidence that the  superconducting state of this material
has odd parity, breaks time reversal symmetry and is fully
gaped.\cite{nelson,kidwingira,Xia,Luke,rmp_pwave,maeno,kapitulnik} One of the
simplest forms of the order parameter which satisfies these requirements is the
chiral p-wave state $\Delta({\bf p}) \sim p_{x}\pm ip_{y}$, which has been
suggested in Ref.~\onlinecite{sigrist}.  It is a two-dimensional analog of
superfluid $^{3}He-A$.\cite{AndersonMorel} Another interesting scenario for the
order parameter was suggested in Ref.~\onlinecite{kivelson}.  Chirality of the
pairing wave function leads to edge states and spontaneous surface currents.
While the quasiparticle tunneling spectroscopy\cite{Kashiwaya,Laube,Liu}
confirmed the existence of the subgap states, the experiments in
Ref.~\onlinecite{Kirtley} did not confirm the existence of the edge
supercurrent. (See Ref.~\onlinecite{Kallin} for a discussion about about
consistency of  the chiral p-wave phase for $Sr_{2}RuO_{4}$.) We think that
electron transport experiments  may clarify the situation.

\begin{figure}[h!]
		\centering \includegraphics[width=0.55\textwidth]{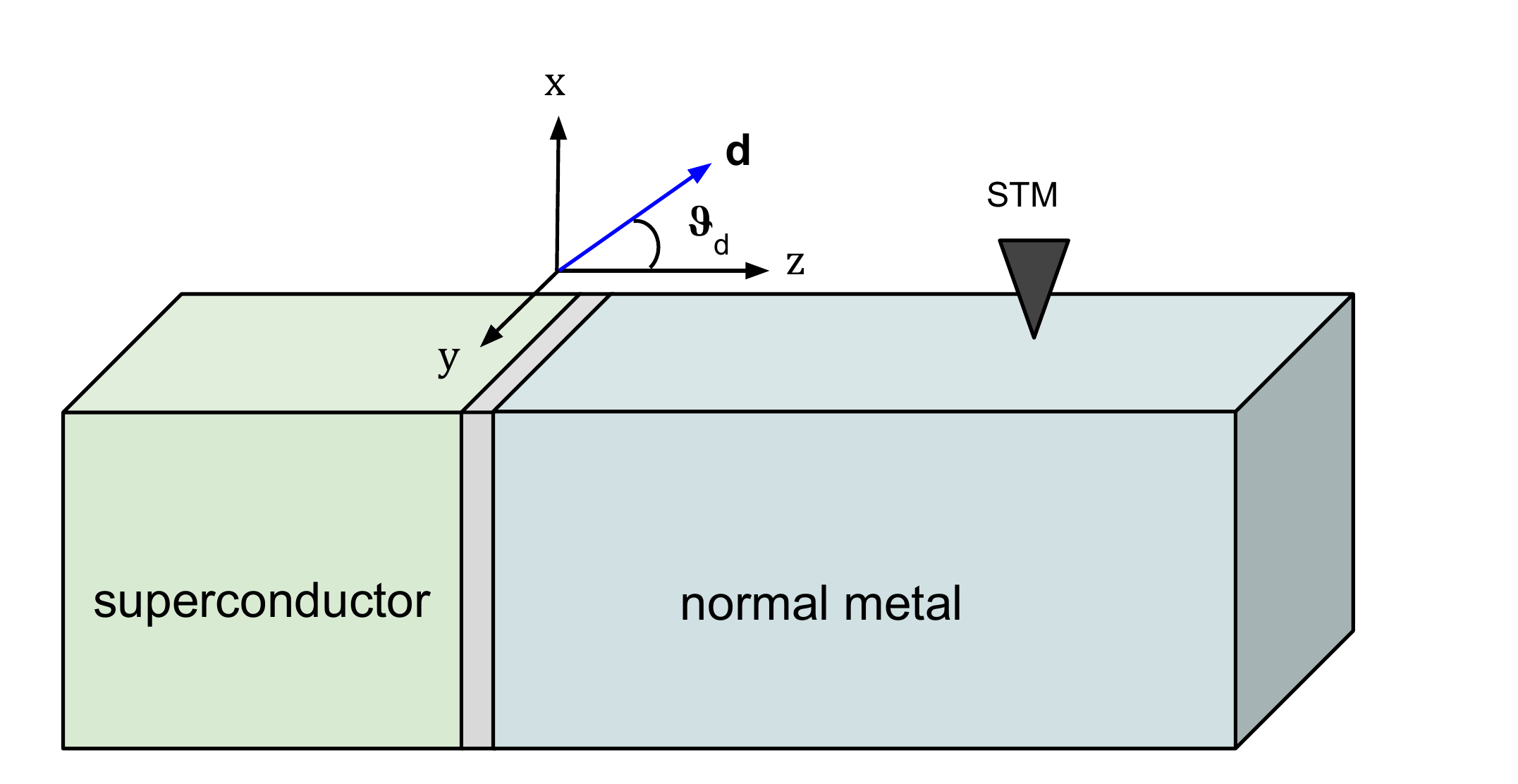}
		\caption{ Schematic representation of the superconductor- insulator-normal
			metal junction. The vector ${\hat z}$ is along the c-axis of crystal, and
			$\vartheta_\mathbf{d}$ denotes the angle between spin vector $\mathbf{d}$
			and  $\hat z$.  The dependence of the voltage inside the normal metal on
			the distance  from the boundary may be measured by a scanning tunneling
			microscope (STM). } \label{fig:geometry}
\end{figure}

In this article we consider a p-wave superconductor-insulator-normal metal
junction in the geometry in which the insulating boundary ($xy$-plane) is
perpendicular to the c-axis of the layered chiral p-wave superconductor, as
shown in Fig. (\ref{fig:geometry}).  Although for simplicity we take the order
parameter in the superconductor in the form\cite{AndersonMorel,mineev}
\begin{equation}\label{eq:delta}
		\hat{\Delta}({\bf n})=\Delta({\bf n})(\mathbf{d}\cdot
		\boldsymbol{\sigma})i \sigma_2, \quad \Delta({\bf n})
		=\Delta_0e^{i\varphi_\mathbf{n}},
\end{equation}
our results also apply to more complicated forms of the order parameter, such as
for example that in Ref.~\onlinecite{kivelson}. Here $\mathbf{n}$ is a unit
vector in the xy-plane, which points along $\mathbf{p}$, and
$\varphi_\mathbf{n}$ is the azimuthal angle characterizing its direction
$\mathbf{n} =(\cos\varphi_\mathbf{n}, \sin \varphi_\mathbf{n})$.

At temperatures well below the gap, tunneling of single electrons from the metal
to the superconductor is forbidden. Thus, similar to the s-wave case, the
resistance of the junction is determined by the tunneling of the electron pairs.
Coherent pair tunneling gives rise to  coherence between electrons and holes
inside the normal metal. Electron-hole coherence in the metal is characterized
by the anomalous Green function. The crucial difference between the s-wave and
the p-wave cases is the following.  In the p-wave case in the absence of
spin-orbit interaction only the p-wave component is induced inside the normal
metal.  The latter is exponentially suppressed at distances larger than $l$ away
from the superconductor-normal metal boundary. As a result, in the diffusive
regime the conductance of the junction is significantly suppressed. In the
presence of the spin-orbit interaction the p-wave order parameter in the
superconductor is partially converted to the s-wave component inside the normal
metal. At low temperatures, the s-wave component  propagates into the metal to
large distances from the boundary. Consequently, it is this component that
determines the low temperature resistance of the system.

We show below that Rashba-type spin-orbit coupling at the boundary between the
normal metal and the p-wave superconductor leads to strong dependence of the
resistance on the direction of the vector ${\bf d}$, which characterizes the
spin structure of the order parameter in Eq.~(\ref{eq:delta}). Since the spin
orientation of the order parameter may be controlled by an external magnetic
field\cite{knight} our predictions may be tested in experiment. Qualitatively,
this dependence may be understood as follows.  In our geometry (with the
$z$-axis parallel to the c-axis of the crystal) the $z$-component of the total
(orbital plus spin) angular momentum, $J_z=L_z + s_z$, is conserved during
tunneling even in the presence of spin-orbit interaction.  Therefore the s-wave
singlet proximity effect in the normal metal is produced only by the pairs with
$J_z=0$ in the p-wave superconductor. Since in our geometry the $z$-component of
the orbital angular momentum in a $p_x+i p_y$ superconductor is $L_z=+1$ we
conclude that only the part of the condensate with $s_z=-1$ induces the s-wave
proximity effect in the normal metal. This condensate fraction corresponds to
the the components of the vector $\mathbf{d}$ lying in the xy-plane.

\section{Kinetic scheme for description of electron transport in p-wave
superconductor-normal metal junctions.}
\label{sec:kinetic_equations}

The conventional description of the electronic transport in superconductors
based on the Boltzmann kinetic equation is valid when all spatial scales in the
problem, including the mean free path $l$, are larger than the characteristic
size of electron wave packets. At low temperatures, $L_{T}\gg l$, this condition
is violated and this approach cannot be used for the description of the effects
mentioned above.

The  set of equations describing the electronic transport in s-wave
superconductors in this situation has been derived in
Ref.~\onlinecite{LarkinOvchinnikov}. Below we review a modification of this
approach for the case where the superconducting part of the junction is a p-wave
superconductor.  The central object of this approach is the matrix Green
function in the Keldysh space
\begin{equation}\label{eq:gorkov_functions}
		\check{G}({\bf x}_1; {\bf x}_2)=\left(
		\begin{array}{cc}
			\hat{G}^R & \hat{G}^K \\
					0     & \hat{G}^A
		\end{array}\right).
\end{equation}
The retarded, advanced and Keldysh Green functions in this equation can be
written in the following form
\begin{eqnarray}
		\hat{G}_{\ell\ell'}^R(\mathbf{x}_1;\mathbf{x}_2)&=&-i\theta(t_1-t_2)
		\langle
			\{\psi_\ell(\mathbf{x}_1),
			\psi^\dagger_{\ell'}(\mathbf{x}_2)\}
		\rangle,\\
		\hat{G}_{\ell\ell'}^A(\mathbf{x}_1;\mathbf{x}_2)&=\!&i\theta(t_2-t_1)
		\langle
			\{\psi_\ell(\mathbf{x}_1),
			\psi^\dagger_{\ell'}(\mathbf{x}_2)\}
		\rangle,\\
		\hat{G}^K_{\ell\ell'}(\mathbf{x}_1;\mathbf{x}_2)&=&-i
		\langle
			[\psi_\ell(\mathbf{x}_1),
			\psi^\dagger_{\ell'}(\mathbf{x}_2)]
	\rangle.
\end{eqnarray}
Here $\mathbf{x}=(\mathbf{r},t)$ denotes the space-time coordinate, and the
indices $\ell,\ell'=1...4$ label the four components of the fermion operator in
the Nambu/spin space; $\psi_1=\psi_\uparrow$, $\psi_2=\psi_\downarrow$,
$\psi_3=\psi_\uparrow^\dagger$, $\psi_4=\psi_\downarrow^\dagger$. Finally, the
anticommutator and the commutator of operators $A$ and $B$ are denoted by
$\{A,B\}$  and $[A,B]$ respectively.

Introducing the new variables,
$\mathbf{x}=(\mathbf{r},t)=(\mathbf{x}_1+\mathbf{x}_2)/2$, and
$\mathbf{x}'=(\mathbf{r}',t')=\mathbf{x}_1-\mathbf{x}_2$,  we can define  the
quasiclassical  Green function  by Fourier transforming $\check{G}({\bf x}_1;
{\bf x}_2)$ with respect to $\mathbf{x}'$ and integrating over
$\xi_\mathbf{p}=\varepsilon_\mathbf{p}-E_F$  as
\begin{equation}\label{eq:quasiclassical_function}
		\check{g}(\mathbf{x},\mathbf{n},\epsilon)=
		\frac{i}{\pi}\int d\xi_\mathbf{p}
		\int d^4x'
		e^{i\epsilon t'-i\mathbf{p}\mathbf{r}'}
		\tau_3
		\check{G}(\mathbf{x}_1;\mathbf{x}_2).
\end{equation}
Here $E_F$ is the Fermi energy, $\varepsilon_\mathbf{p}$ is the electron energy
spectrum, and $\mathbf{n}$ is a unit vector labeling a location on the Fermi
surface (for example, for a spherical Fermi surface it can be chosen as
$\mathbf{n}=\mathbf{p}/|\mathbf{p}|$), and  $\tau_3$ is the third Pauli matrix.
In this paper, we will denote the  Pauli matrices in the Nambu space by $\tau_i$, and
the Pauli matrices in spin space by $\sigma_i$. The Keldysh space structure of the Green
functions will be indicated explicitly when necessary.

The quasiclassical Green's function (\ref{eq:quasiclassical_function}) satisfies
the normalization condition
\begin{equation}
		\label{eq:norm}
		\check{g}\check{g}=1,
\end{equation}
which can be spelled out in terms of components in the Keldysh space as,
\begin{eqnarray}
		\hat{g}^{(R,A)}\hat{g}^{(R,A)}=1\label{eq:norm_E},\\
		\hat{g}^R\hat{g}^K+\hat{g}^K\hat{g}^A=0.\label{eq:norm_NE}
\end{eqnarray}
The normalization condition Eq.~(\ref{eq:norm_NE}) is satisfied for any Keldysh function
of the form
\begin{equation}\label{eq:keldysh}
		\hat{g}^K=\hat{g}^R\hat{h}-\hat{h}\hat{g}^A.
\end{equation}
The matrix $\hat{h}$ may be parameterized
as\cite{LarkinOvchinnikov}
\begin{equation}\label{eq:LO_parametrization}
		\hat{h}=f_0\hat{\tau}_0+f_1\hat{\tau}_3.
\end{equation}
Here $f_0$ and $f_1$ are respectively the odd and even in
$\epsilon$ parts of the distribution function (see Ref.~\onlinecite{shelankov}
for an alternative treatment).

In this paper we only consider stationary situations. In this case the Green
functions depend on the energy $\epsilon$ but not on the time $t$. If $k_{F}l\gg
1$ the Gorkov equation for the Green function in Eq.~(\ref{eq:gorkov_functions})
can be reduced to the quasi-classical Eilenberger equations for the Green
functions defined in Eq.~(\ref{eq:quasiclassical_function})
\cite{LarkinOvchinnikov}
\begin{equation}
		i\mathbf{v}_F\cdot\boldsymbol{\nabla}\check{g}+
		\left[
				\epsilon\check{\tau}_3-\check{\Delta}(\mathbf{n})-\check{\Sigma},
				\check{g}
		\right]=0. \label{eq:EilenbergerEq}
\end{equation}
Here  $\check{\Sigma}$ is the self energy associated with impurity scattering.
In the Born approximation $\check{\Sigma}=-i\langle\check{g}\rangle/2\tau_e$,
where $\langle \ldots \rangle$ denotes average over the solid angle in momentum
space and $\tau_e$ is the elastic mean free time. The only difference of
Eq.~(\ref{eq:EilenbergerEq}) from the conventional s-wave superconductor case is in
the form Eq.~(\ref{eq:delta}) for the order parameter.

We neglect the electron-electron interactions in the normal metal. As a result,
in our approximation the order parameter vanishes inside the normal metal.  This
yields the following equations for the retarded, advanced and Keldysh  Green
functions:
\begin{eqnarray}
		i\mathbf{v}_F\cdot\boldsymbol{\nabla}\hat{g}^{R,A}+
		\epsilon[\hat{\tau}_3,\hat{g}^{R,A}]
		&=&[\hat{\Sigma}^{R,A},\hat{g}^{R,A}],  \\
		\label{eq:keldysh2}
		i\mathbf{v}_F\cdot\boldsymbol{\nabla}\hat{g}^{K} + \epsilon[\tau_3,\hat{g}^{K}]
		&=&\hat{\Sigma}^R\hat{g}^K + \hat{\Sigma}^K\hat{g}^A \nonumber \\
	&& - \hat{g}^R\hat{\Sigma}^K -\hat{g}^K\hat{\Sigma}^A.
\end{eqnarray}
Multiplying Eq. (\ref{eq:keldysh2}) with $\tau_3$ and $\tau_0$ and taking the
trace, and using the fact that $\mathrm{Tr} (\hat{g}^R-\hat{g}^A)=0$,  one
obtains the following equations for $f_1$ and $f_0$
\begin{widetext}
	\begin{eqnarray}\label{eq:f_1_transport}
		\mathrm{Tr}\left[\hat{\beta}\right]\mathbf{v}_F\cdot\boldsymbol{\nabla} f_1
		&=&
		-\frac{1}{2\tau_e}f_0\mathrm{Tr}
		\left(
			\langle \hat{\alpha}\rangle\hat{\alpha}-
			[\langle \hat{g}^R\rangle,\hat{g}^R]+
			[\langle \hat{g}^A\rangle,\hat{g}^A]
		\right)
		+\frac{1}{2\tau_e}\mathrm{Tr}
		\left[
			\langle \hat{\alpha} f_0\rangle\hat{\alpha}
		\right]\nonumber\\
		&-&\frac{1}{2\tau_e}f_1\mathrm{Tr}
		\left(
			\langle \hat{\alpha}\rangle\hat\beta-
			[\langle \hat g^R\rangle,\hat g^R]\hat\tau_3+
			\hat\tau_3[\langle \hat g^A\rangle,\hat g^A]
		\right)
		+\frac{1}{2\tau_e}\mathrm{Tr}
		\left[
			\langle \hat\beta f_1\rangle\hat\alpha
		\right],\\
		\label{eq:f_0_transport}
		\mathrm{Tr}\left[\hat\beta\right]\mathbf{v}_F\cdot\boldsymbol{\nabla} f_0
		&=&
		-\frac{1}{2\tau_e}f_0\mathrm{Tr}
		\left(
			\langle \hat\tau_3\hat\beta\hat\tau_3\rangle\hat\alpha-
			[\langle \hat g^R\rangle,\hat g^R]\hat\tau_3+
			\hat\tau_3[\langle \hat g^A\rangle,\hat g^A]
		\right)
		+\frac{1}{2\tau_e}\mathrm{Tr}
		\left[
			\langle\hat \alpha f_0\rangle\hat\tau_3\hat\beta\hat\tau_3
		\right]\nonumber\\
		&-&\frac{1}{2\tau_e}f_1\mathrm{Tr}
		\left(
			\langle\hat g^R\rangle\hat\beta\hat\tau_3
			-\hat\tau_3\hat\beta\langle\hat g^A\rangle-
			[\langle\hat g^R\rangle,\hat g^R]+
			[\langle\hat g^A\rangle,\hat g^A]
		\right)
		+\frac{1}{2\tau_e}\mathrm{Tr}
		\left[
			\langle\hat \beta f_1\rangle\hat\tau_3\hat\beta\hat\tau_3
		\right].
\end{eqnarray}
\end{widetext}
Here we defined $\hat \alpha= \hat g^R- \hat g^A$ and $\hat\beta=\hat
g^R\hat\tau_3-\hat\tau_3\hat g^A$.

The gauge-invariant potential  and the electric current can be expressed
in terms of quasiclassical Keldysh green functions as,
\begin{eqnarray}\label{eq:Phi_def}
		\Phi(\mathbf{r})
		&=&\frac{1}{4e}\int d\epsilon\int d^2\mathbf{n}
		\mathrm{Tr}\{\hat{g}^K(\mathbf{r},\mathbf{n},\epsilon)\}\\
		J(\mathbf{r})
		&=&	-\frac{e\nu_0}{4}\int d\epsilon\int d^2\mathbf{n}
		\mathbf{v}_F\mathrm{Tr}
		\{\hat{\tau}_3\hat{g}^K(\mathbf{r},\mathbf{n},\epsilon)\}. \label{eq:J_def}
\end{eqnarray}
Here the integral over $\mathbf{n}$ denotes averaging over the Fermi surface,
$d^2 \mathbf{n}=d \Omega_\mathbf{n}/4\pi$.

We discuss the boundary conditions for the quasiclassical transport equations
(\ref{eq:EilenbergerEq}) - (\ref{eq:f_0_transport})
in Sec. \ref{sec:boundary_conditions}.

\subsection{Boundary conditions for p-wave superconductor-normal metal interface}
\label{sec:boundary_conditions}

The p-wave superconductivity is destroyed by elastic scattering processes when
$l< \xi_0$, where $\xi_0$ is the zero temperature coherence length in a clean
superconductor. Therefore we consider the case where the p-wave superconductor
is relatively clean and $l\gg \xi$.  For the same reason the p-wave proximity
effect is exponentially suppressed in the metal at distances larger than $l$
from the boundary.  On the other hand, in a spatially inhomogeneous system in
the presence of spin-orbit interaction the p- and s- wave components of the
anomalous Green functions are mixed. At low temperatures, the s-wave component
induced by spin-orbit coupling extends into the metal to distances much larger
than $l$, and determines the low temperature transport properties of the
junction. Therefore spin-orbit coupling plays a crucial role in low temperature
electron transport in normal metal--p-wave superconductor junctions.

Though our results have a general character, in this article  we
assume that a Rashba type spin orbit coupling is present only at the boundary.
 The corresponding potential
energy at the boundary may be modeled by the form $V=
(u_0\sigma_0+u_1\hat{z}\times\mathbf{p}_\parallel\cdot\boldsymbol{\sigma})\delta(z)$,
where $\mathbf{p}_\parallel$ is the component of the electron
momentum parallel to the boundary,  and $\hat{z}$
is the unit vector normal to the boundary. We assume that $u_1\ll u_0$, and consider a
disorder free boundary, so  that  $\mathbf{p}_\parallel$  is conserved.

\begin{figure}[h!]
	\centering \includegraphics[width=0.5\textwidth]{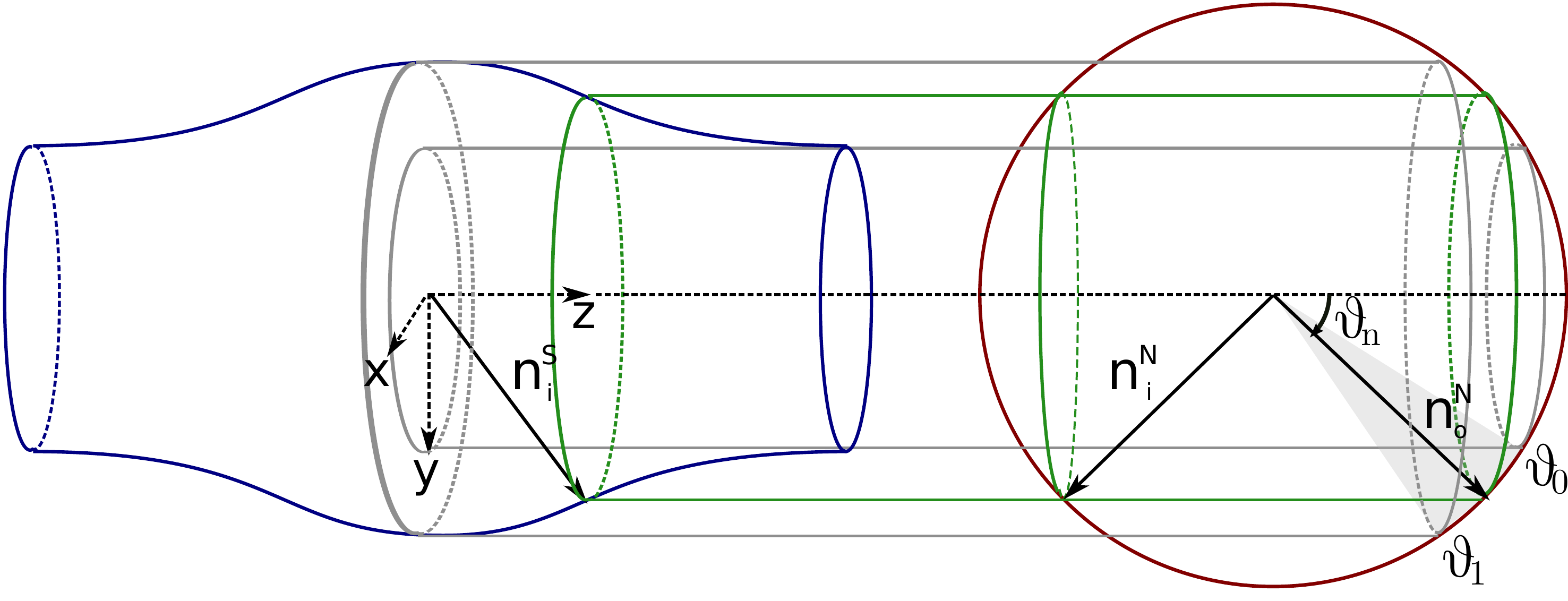}
	\caption{Fermi surface topologies of the superconductor (corrugated cylinder
		at left) and the normal metal (sphere at right). The vectors $\mathbf{n}_i$
		and  $\mathbf{n}_o$ correspond to respectively incident and outgoing waves.
		The superscripts $N$ and $S$ denote the superconductor and the normal metal
		sides of the insulating barrier. The vectors $\mathbf{n}^N$ and
		$\mathbf{n}^S$ correspond to the same parallel momentum, as shown by the
		green lines.  The momentum domain where tunneling is possible is bounded by
		the angles $\vartheta_0$ and $\vartheta_1$. These angles define the
		integration limits in Eqs.~(\ref{eq:matrix_current_boundary}) and
		(\ref{eq:t2}).  } \label{fig:momentum_boundary}
\end{figure}

The boundary conditions for quasiclassical Green functions in superconductors
were obtained in Refs.~\onlinecite{zaitsev,millis,kuprianov}. In the case of a
spin active boundary~\cite{millis} they may be expressed in terms of the
$\mathbf{p}_\parallel$-dependent  scattering matrix of the insulating barrier.
The latter relates the spinor amplitudes of the outgoing ($\psi_o$) and incident
($\psi_i$) electron waves,
\begin{equation}\label{eq:S_matrix}
	\left(
		\begin{array}{c}
			\psi_o^S \\
			\psi_o^N \\
		\end{array}
	\right)
	=\left(
		\begin{array}{cc}
			S_{11} & S_{12} \\
			S_{21} & S_{22} \\
		\end{array}
	\right) \left(
		\begin{array}{c}
			\psi_i^S \\
			\psi_i^N \\
		\end{array}
	\right).
\end{equation}
Here the superscripts $N$ and $S$ denote respectively the normal metal and the
superconductor side of the barrier.  The presence of spin-orbit interaction at
the boundary results in a spin-dependent transmission amplitude $S_{12}$, which
may be written in the form
\begin{eqnarray}\label{eq:tunneling_so}
	S_{12}&=&
	t_0+ t_s  \gamma (\varphi_\mathbf{n}),  \\
	\label{eq:gamma_n_def}
	\gamma(\varphi_\mathbf{n})&=& \cos \varphi_\mathbf{n} \sigma_y - \sin \varphi_\mathbf{n} \sigma_x .
\end{eqnarray}
Here we introduced the azimuthal angle $\varphi_\mathbf{n}$ as
$\mathbf{p}_\parallel=|\mathbf{p}_\parallel| (\hat x \cos \varphi_\mathbf{n} +
\hat y \sin \varphi_\mathbf{n})$. The spin-dependent  and spin-independent
transmission amplitudes  $t_s$ and $t_0$, are scalar functions of
$|\mathbf{p}_\parallel|$. To lowest order in the transmission amplitude, the
boundary condition for the quasiclassical Green functions may be written
as~\cite{millis}
\begin{eqnarray}	
		\check{g}(\mathbf{r}^{N},\mathbf{n}^N_{{o}})
		&=&
		-\frac{1}{2}
		\left[
				\check S_{21}
				\left(
						\check{g}(\mathbf{r}^{S},\mathbf{n}^S_{{i}})-1
				\right)
				\check{S}^\dagger_{21},
				\check{g}(\mathbf{r}^{N},\mathbf{n}^N_{{o}})
		\right]   \nonumber\\
		&&+
		\check{S}_{22}
		\check{g}(\mathbf{r}^{N},\mathbf{n}^N_{{i}})
		\check{S}^{-1}_{22},\label{eq:boundary-millis1}\\	
		\check{g}(\mathbf{r}^{S},\mathbf{n}^S_{{i}})
		&=&
		-\frac{1}{2}
		\left[
				\check{g}(\mathbf{r}^{S},\mathbf{n}^S_{{i}}),
				\check S_{21}^\dagger
				\left(
						\check{g}(\mathbf{r}^{N},\mathbf{n}^N_{{o}})-1
				\right)
				\check{S}_{21}
		\right]   \nonumber\\
		&&+
		\check{S}^{-1}_{11}
		\check{g}(\mathbf{r}^{S},\mathbf{n}^S_{{o}})	
		\check{S}_{11}\label{eq:boundary-millis2}.
\end{eqnarray}
Here $\mathbf{n}_{i,o}^S$ and $\mathbf{n}_{i,o}^N$ are the unit vectors
indicating positions on the Fermi surface in the superconductor ($S$) and the
normal metal ($N$) for the incident ($i$) and outgoing ($o$) waves. By momentum
conservation they correspond to the same $\mathbf{p}_\parallel$ and thus are
characterized by the same azimuthal angle $\varphi_\mathbf{n}$.  For simplicity
we assume that Fermi surface in the superconductor to be a corrugated cylinder
with the symmetry axis along $\hat z$, and that in the normal metal to be a sphere.
The Fermi surface points corresponding to the incident and reflected waves are
illustrated in Fig.~\ref{fig:momentum_boundary}.  The coordinates $\mathbf{r}^N$
and $\mathbf{r}^S$ correspond respectively to the normal metal - and the
superconductor- sides of the insulating boundary. For brevity the obvious
$\epsilon$ dependence of Green functions has been dropped. Finally the matrices
$\check S_{\alpha\beta}$ are defined following Ref.~\onlinecite{millis} as
\begin{equation}\label{eq:S_hat_def}
	\check S_{\alpha\beta}=
	S_{\alpha\beta}(\mathbf{p}_\parallel)\frac{1+\tau_3}{2}+
	S_{\beta\alpha}(-\mathbf{p}_\parallel)^{T}\frac{1-\tau_3}{2},
\end{equation}
where  $S_{\alpha\beta}$ is defined in Eq.~(\ref{eq:tunneling_so})  and the
superscript ${T}$ denotes the matrix transposition in the spin space.
At weak tunneling we may
approximate $\check{S}_{11}\approx\check{S}_{22}\approx 1$, and
\begin{equation}
	\check{S}_{12}=t_0\check{1}+t_s \check{\gamma}.
\end{equation}
Here we introduced
\begin{eqnarray} \label{eq:hat_gamma_def}
	\check{\gamma}=
	\left(
		\begin{array}{cc}
			\hat\gamma   & 0  \\
			0 	& \hat\gamma  \\
		\end{array}
	\right),\quad
	\hat{\gamma}=
	\left(
		\begin{array}{cc}
			\gamma(\varphi_\mathbf{n})   & 0  \\
			0 & -\gamma(\varphi_\mathbf{n})^{T}  \\
		\end{array}
	\right).
\end{eqnarray}
with
$\gamma(\varphi_\mathbf{n})$ defined in Eq.~(\ref{eq:gamma_n_def}).

For the purpose of studying electron transport at low temperatures, $T\ll
\Delta$, we only need the Green functions with energies $\epsilon$ well below
the gap $\Delta$. The Green functions inside the
superconductor are practically unaffected by tunneling. Therefore,
the boundary condition for the normal metal Green function is given by
Eq.~(\ref{eq:boundary-millis1}), where the superconductor Green functions are
replaced by their value in the bulk. Since latter do not depend on $p_z$ we have
$\check{g}(\mathbf{r}^S,\mathbf{n}_i^S)= \check{g}(\mathbf{r}^S,\mathbf{n}_o^S)\equiv \check{g}(\mathbf{r}^S,\mathbf{n}^S)$.
It is useful to define symmetric and antisymmetric Green functions as
\citep{zaitsev,kuprianov}
\begin{equation}\label{eq:g_sym_antisym}
	\check{g}_{s,a}(\mathbf{r},\mathbf{n})
	=\frac{1}{2}[\check{g}(\mathbf{r},\mathbf{n}_i)
	\pm\check{g}(\mathbf{r},\mathbf{n}_o)]
\end{equation}
With this notation Eq.~(\ref{eq:boundary-millis1})
may be written as follows:
\begin{eqnarray}\label{eq:spinboundary}
	\check{g}_a(\mathbf{r}^N,\mathbf{n}^N)&=& -\frac{t_0^2}{4} \left[
		\check{g}(\mathbf{r}^S,\mathbf{n}^S),
		\check{g}_s(\mathbf{r}^N,\mathbf{n}^N) \right]\nonumber\\ &&
	-\frac{t_0t_s}{4} \left[ \left\{ \check{\gamma},
			\check{g}(\mathbf{r}^S,\mathbf{n}^S) \right\},
		\check{g}_s(\mathbf{r}^N,\mathbf{n}^N) \right]\nonumber\\ &&
	+\frac{t_0t_s}{2} \left[ \check{\gamma},
		\check{g}_s(\mathbf{r}^N,\mathbf{n}^N) \right].
\end{eqnarray}
Here we for simplicity assume that due to weakness of spin-orbit coupling the
electron tunneling amplitude with spin flip is smaller than that without, $t_s
\ll t_0 \ll 1$. The first term in Eq.~(\ref{eq:spinboundary}) arises from the
spin-conserving tunneling and coincides with that in Ref.~\onlinecite{zaitsev}
at small transparency.  This term dominates electron transport properties of the
junction in the high temperature regime.  The second term comes from the spin
orbit coupling. Although it is smaller than the first one, it generates the s-wave
component proximity effect in the normal metal and thus determines the electron
transport at low temperatures. Finally, the last term  is odd in the parallel
momentum. Therefore it vanishes upon averaging over the Fermi surface and does not contribute to electron transport in the diffusive regime.

\subsection{Kinetic scheme in the diffusive regime}

In the low temperature regime, $T \ll v_F/l$, the proximity effect extends to
distances of order $L_T=\sqrt{D/T} \gg l $ into the normal metal (here $D$ is
the electron diffusion constant). At such length scales the transport properties
of the junction may be described in terms of the Usadel Green functions
$\check{G}(\mathbf{r})$. The latter correspond to coincident coordinates of the
electron operators in Eq.~(\ref{eq:gorkov_functions}), $\mathbf{r}=\mathbf{r}'$,
and may be expressed in terms of the Eilenberger Green functions
(\ref{eq:quasiclassical_function}) by averaging them over the Fermi surface
\begin{eqnarray}\label{eq:G_Usadel_def}
	\check{G}(\mathbf{r})
	=\int d^2\mathbf{n}\,
	\check{g}(\mathbf{r},\mathbf{n}), \quad d^2\mathbf{n}
	=\frac{1}{4\pi}\, d\cos\vartheta_\mathbf{n} d\varphi_\mathbf{n}.
\end{eqnarray}
where the polar and azimuthal angles $\vartheta_\mathbf{n}$ and
$\varphi_\mathbf{n}$ characterize the unit vector $\mathbf{n}=(n_x, n_y, n_z)=
(\sin \vartheta_\mathbf{n} \cos \varphi_\mathbf{n}, \sin \vartheta_\mathbf{n}
\sin \varphi_\mathbf{n}, \cos \vartheta_\mathbf{n})$.

We neglect the spin-orbit interaction in the normal metal and assume that the
electrons in the normal lead are not spin polarized. The triplet component of
the anomalous Green function is exponentially suppressed at distances larger
than $l$ from the boundary with the superconductor. The singlet component, on
the other hand, survives even at distances much larger than $l$. Therefore it
dominates the electron transport in the junction at low temperatures. Below we
focus on the singlet component of the Usadel Green function, $\hat G
(\mathbf{r})$, which is a $4\times 4$ matrix in the Keldysh and Nambu space. Its
various components $\alpha=R, A, K$ in the Keldysh space have the following form
\begin{equation}\label{eq:G_singlet}
	\hat{G}^{\alpha}(\mathbf{r})=
	\left(
		\begin{array}{ccc}
			G^{\alpha} & -iF^{\alpha} \\
			i\tilde F^{\alpha}  & -\tilde G^{\alpha}
		\end{array}
	\right).
\end{equation}
The corresponding spin structure of the full $8\times 8$ Green function in
Eq.~(\ref{eq:G_Usadel_def}) is given by
\begin{equation}
	\check{G}^{\alpha}(\mathbf{r})=
	\left(
		\begin{array}{ccc}
			G^{\alpha} \sigma_0 & -iF^{\alpha} i\sigma_2\\
			i\tilde F^{\alpha} i\sigma_2 & -\tilde G^{\alpha} \sigma_0
		\end{array}
	\right).
\end{equation}

At length scales greater than $l$ the singlet component of the Usadel Green
function satisfies the differential equation
\begin{equation}\label{eq:usadel}
	D\boldsymbol{\nabla} \cdot
	\left[
		\hat{G}(\mathbf{r})
		\boldsymbol{\nabla}\hat{G}(\mathbf{r})
	\right]
	+i\epsilon
	\left[
		\hat{\tau}_3,
		\hat{G}(\mathbf{r})
	\right]=0.
\end{equation}
Expanding in the Keldysh space, this equation gives
\begin{eqnarray}
	D\boldsymbol{\nabla} \cdot
	\left(
		\hat{G}^{(R,A)}
		\boldsymbol{\nabla}\hat{G}^{(R,A)}
	\right)
	+i\epsilon[\hat{\tau}_3,\hat{G}^{(R,A)}]=0,
	\label{eq:usadel_E}\\
	D\boldsymbol{\nabla} \cdot
	\left(
		\hat{G}^R
		\boldsymbol{\nabla}\hat{G}^K+
		\hat{G}^K
		\boldsymbol{\nabla}\hat{G}^A
	\right)
	+i\epsilon[\hat{\tau}_3,\hat{G}^K]=0.
	\label{eq:usadel_NE}
\end{eqnarray}
The first  equation (\ref{eq:usadel_E}) is the Usadel equation, which describes
the equilibrium properties of the system. The second equation
(\ref{eq:usadel_NE}) for Keldysh component describes the non-equilibrium
properties. The Usadel Green function satisfies the normalization conditions
(\ref{eq:norm_E}) and (\ref{eq:norm_NE}). The condition (\ref{eq:norm_NE}) is
satisfied by any matrix of the form (\ref{eq:keldysh}). In the normal metal the
matrix $\hat h$ may be expressed in terms of the symmetric and antisymmetric
distribution functions $f_0$ and $f_1$ using
Eq.~(\ref{eq:LO_parametrization}).~\cite{LarkinOvchinnikov}

In a normal metal in contact with a single superconducting lead,
Eq.~(\ref{eq:usadel_NE}) can be used to obtain following equations for
distribution functions by using Eqs.~(\ref{eq:keldysh}) and (\ref{eq:LO_parametrization}):
\begin{eqnarray}
	\boldsymbol{\nabla} \cdot
	\left(
		\mathrm{Tr}
		\left[
			1-\hat{G}^R(\mathbf{r})\hat{G}^A(\mathbf{r})
		\right]
		\boldsymbol{\nabla}f_0(\mathbf{r},\epsilon)
	\right)&=&0,
	\label{eq:even-distribution}\\
	\boldsymbol{\nabla} \cdot
	\left(
		\mathrm{Tr}
		\left[
			1-\tau_3\hat{G}^R(\mathbf{r})\tau_3\hat{G}^A(\mathbf{r})
		\right]
		\boldsymbol{\nabla}f_1(\mathbf{r},\epsilon)
	\right)&=&0.
	\label{eq:odd-distribution}
\end{eqnarray}

The expressions for the density of states, electrochemical potential and current
density in terms of the Usadel Green functions are
\begin{eqnarray}\label{eq:dos}
		\nu(\mathbf{r},\epsilon)
		&=&\nu_0\mathrm{Re}\left\{ G^R(\mathbf{r},\epsilon)\right\},\\
		\Phi(\mathbf{r})
		&=&\frac{1}{e\nu_0}\int d\epsilon \nu(\mathbf{r},\epsilon)f_1(\mathbf{r},\epsilon),
		\label{eq:potential1}\\
		J(\mathbf{r})
		&=&e\nu_0D \int d\epsilon
		\Pi(\mathbf{r},\epsilon)\boldsymbol{\nabla} f_1(\mathbf{r},\epsilon).
		\label{eq:current1}
\end{eqnarray}
Here $\Pi(\mathbf{r},\epsilon)
=1+|G^R(\mathbf{r},\epsilon)|^2+|F^R(\mathbf{r},\epsilon)|^2$, $\nu_{0}=
mp_F/\pi^2$ is the density of states of the normal metal in the absence of the
proximity effect.

Using  Eq.~(\ref{eq:norm_E}) one can write the retarded Usadel Green function in
terms of the complex angles $\theta(\mathbf{r})$ and
$\chi(\mathbf{r})$ as
\begin{equation}\label{eq:G^R_theta}
	\hat G^R(\mathbf{r})=
	\left(
		\begin{array}{cc}
			\cos\theta(\mathbf{r}) & -i \sin\theta(\mathbf{r}) e^{i \chi(\mathbf{r})} \\
			i \sin\theta(\mathbf{r}) e^{-i \chi(\mathbf{r})} & -\cos\theta(\mathbf{r}) \\
		\end{array}
	\right).
\end{equation}
The corresponding parametrization for advanced  Green function can be obtained
by using $\hat G^A(\mathbf{r})=-\tau_3\left[\hat G^R(\mathbf{r})\right]^\dagger\tau_3$.

For the system of interest, where the normal metal is connected to a single
superconductor the phase $\chi(\mathbf{r})$ is independent of coordinates and is
set by the phase of the order parameter in the superconductor.  In this case
($\boldsymbol{\nabla}\chi(\mathbf{r})=0 $) the Usadel equation in
(\ref{eq:usadel_E}) reduces to the following second order differential equation
for the complex function $\theta(\mathbf{r})$:
\begin{equation}
	\frac{D}{2}\boldsymbol{\nabla}^{2}\theta(\mathbf{r})
	+i\epsilon\sin\theta(\mathbf{r})=0\label{eq:sinegordon},
\end{equation}
which is the well known sine-Gordon equation.

The equations for the distribution functions in (\ref{eq:even-distribution}) and
(\ref{eq:odd-distribution}) take the following forms in this parametrization:
\begin{eqnarray}
	D\boldsymbol{\nabla} \cdot
	\left(
		\cos^2\theta_R(\mathbf{r})\boldsymbol{\nabla} f_0(\mathbf{r})
	\right)
	&=& 0 \label{eq:odd-distribution-theta},\\
	D\boldsymbol{\nabla} \cdot
	\left(
		\cosh^2\theta_{I}(\mathbf{r}) \boldsymbol{\nabla}f_1(\mathbf{r})
	\right)
	&=& 0 \label{eq:even-distribution-theta}.
\end{eqnarray}
Here we introduced the real and imaginary parts of $\theta(\mathbf{r})=
\theta_R(\mathbf{r})+i\theta_I(\mathbf{r})$.

Finally,   using Eqs.~(\ref{eq:potential1}), (\ref{eq:current1}), and
(\ref{eq:G^R_theta}) we get the following expressions for the electric current
and potential
\begin{eqnarray}
	J_n(\mathbf{r})
	&=&eD\nu_0
	\int d\epsilon\cosh^2\theta_I(\mathbf{r})\boldsymbol{\nabla} f_1(\mathbf{r})
	\label{eq:current}\\
	\Phi(\mathbf{r})
	&=&\frac{1}{e}
	\int d\epsilon\cos\theta_R(\mathbf{r})\cosh\theta_I(\mathbf{r})f_1(\mathbf{r})
	\label{eq:potential}.
\end{eqnarray}
Below we will be interested only in linear in the external electric field
effects, in which case $f_{0}=\tanh(\epsilon/2T)$, has its equilibrium form.

The equations (\ref{eq:usadel_E}-\ref{eq:usadel_NE}) or
(\ref{eq:sinegordon}-\ref{eq:even-distribution-theta}) must be supplemented with
the boundary conditions. In Sec. \ref{sec:Usadel_bc} we obtain such conditions
for a boundary between the normal metal and the p-wave superconductor in the
geometry of our device.

\subsubsection{Diffusive Boundary Conditions in the Vertical Geometry }
\label{sec:Usadel_bc}

The boundary conditions for the Usadel Green function $\hat G (\mathbf{r})$ may
be found by solving the Eilenberger equations (\ref{eq:EilenbergerEq}) with
boundary conditions (\ref{eq:spinboundary}) at distances of the order of the
mean free path $l$ from the boundary. This can be done using the method of
Ref.~\onlinecite{kuprianov}. A key observation is that the Eilenberger equations
(in which one may set $\epsilon \to 0$ for distances less than the mean free
path from the boundary) conserve the matrix current normal to the boundary,
\begin{equation}\label{eq:matrix_current}
	\check j (\mathbf{r})
	=\int d^2\mathbf{n} \check{g}({\bf r},\mathbf{n})\mathbf{v}_F \cdot \hat z
	=v_F\int' d^2\mathbf{n} \check{g}_a({\bf r},\mathbf{n}) \mathbf{n} \cdot \hat z .
\end{equation}
The prime in the second expression indicates the fact that the integral must be
taken over half the Fermi surface, $\mathbf{n}\cdot \hat z \geq 0$.

At weak tunneling the singlet component $\hat j$ of the matrix current at the
boundary may be expressed in terms of the Usadel Green function $\hat G
(\mathbf{r})$ as\cite{kuprianov}
\begin{equation}\label{eq:matrix_current_Usadel}
	\hat j(\mathbf{r}^N)=  D\hat{G}(\mathbf{r})\hat z\cdot
	\boldsymbol\nabla\hat{G}(\mathbf{r})|_{\mathbf{r}=\mathbf{r}^N}.
\end{equation}
On the other hand, the matrix current may be evaluated by multiplying
Eq.~(\ref{eq:spinboundary}) with $v_F \mathbf{n}^N \cdot \hat z$ and integrating
the result over half the Fermi surface, $\mathbf{n}^N\cdot \hat z \geq 0$.  In
doing so it is important to keep in mind that at weak tunneling the symmetric
part of Green function in the normal metal is independent of $\mathbf{n}^N$,
$\check{g}_s(\mathbf{r}^N,\mathbf{n}^N)=\check{G}(\mathbf{r}^N)$, and that  the
superconductor Green function $\check{g}_s(\mathbf{r}^S,\mathbf{n}^S)$ may be
replaced by its bulk value at $\epsilon=0$. The latter is given by
\begin{eqnarray}
	\check{g}(\mathbf{n})=-
	\left[
		\begin{array}{cc}
			0 &	e^{i(\varphi_\mathbf{n}+\chi_0)}
			\mathbf{d}\cdot\boldsymbol{\sigma}i \sigma_2\\
			e^{-i(\varphi_\mathbf{n}+\chi_0)}i\sigma_2
			\mathbf{d}^*\cdot\boldsymbol{\sigma}
			& 0
		\end{array}
	\right].
	\label{eq:greenfunction_p}
\end{eqnarray}
Here we used  Eq.~(\ref{eq:delta}). We consider unitary states,
\mbox{$\mathbf{d}\times\mathbf{d}^*=0$}, and parameterize the vector $\mathbf{d}$ by an overall phase
$\chi_0$ and the spherical angles
$\vartheta_\mathbf{d}$, and $\varphi_\mathbf{d}$ as,
\begin{equation}\label{eq:d_angles}
	\mathbf{d}^T
	=e^{i\chi_0}
	(\sin\vartheta_\mathbf{d}\cos\varphi_\mathbf{d},
	\sin\vartheta_\mathbf{d} \sin\varphi_\mathbf{d},
	\cos\vartheta_\mathbf{d}).
\end{equation}

It is easy to see that only the second term in the right hand side of
Eq.~(\ref{eq:spinboundary}) contributes to the matrix current. The contributions
of the other two terms vanish upon the integration over $\mathbf{n}$ because
both $\check\gamma$ and the superconductor Green function
$\check{g}(\mathbf{r}^S,\mathbf{n}^S)$ depend on the azimuthal angle
$\varphi_\mathbf{n}$ as $e^{\pm i \varphi_\mathbf{n}}$, see
Eqs.~(\ref{eq:gamma_n_def}), (\ref{eq:hat_gamma_def}) and
(\ref{eq:greenfunction_p}). We thus obtain
\begin{equation}\label{eq:matrix_current_boundary}
	\check j (\mathbf{r})= -\frac{ v_F}{4}\int_{\vartheta_0}^{\vartheta_1} \frac{d \cos \vartheta_\mathbf{n}}{2} t_st_0 \hat z\cdot\mathbf{n}
	\left[ \check{ \bar G}(\mathbf{r}^S)
		,
		\check{G}(\mathbf{r}^N).
	\right]
\end{equation}
Here $v_F$ is Fermi velocity in the normal metal, and the integration limits $\vartheta_0$ and $\vartheta_1$ define the domain where tunneling is possible. This domain corresponds to the projection of the corrugated cylindrical Fermi surface in the superconductor to the Fermi
surface in the metal,  see  Fig.~\ref{fig:momentum_boundary}. Finally,
$\check{\bar G}(\mathbf{r}^S)$ is given by
\begin{eqnarray}
	\check{\bar G}(\mathbf{r}^S)
	&\equiv&
	\int \frac{d\varphi_\mathbf{p}^S}{2\pi}
	\left\{
		\check{g}(\mathbf{r}^S,\mathbf{n}^S_i),
		\check{\gamma}
	\right\}\nonumber\\
	&=&
	\left[
		\begin{array}{cc}
			0 & e^{i(\varphi_\mathbf{d}+\chi_0)}i\sigma_2 \\
			e^{-i(\varphi_\mathbf{d}+\chi_0)}i\sigma_2 & 0
		\end{array}
	\right].
	\label{eq:eff_greenfunction}
\end{eqnarray}
Comparing Eqs.~(\ref{eq:matrix_current_Usadel}) and
(\ref{eq:matrix_current_boundary}) we obtain the following boundary condition
for the Usadel Green function,
\begin{equation}\label{eq:diffusivespinboundary2}
	D\check{G}(\mathbf{r})
	\partial_z\check{G}(\mathbf{r})|_{\mathbf{r}=\mathbf{r}^N}
	=t
	\left[
		\check{G}(\mathbf{r}^N),
		\check{\bar G}(\mathbf{r}^S)
	\right],
\end{equation}
where
\begin{equation}
	\label{eq:t2}
	t=\frac{1}{4}|\sin\vartheta_\mathbf{d}|
	\int_{\vartheta_0}^{\vartheta_1}\frac{d\cos\vartheta_\mathbf{n}}{2}(t_s
	t_0 v_{F}\cos\vartheta_\mathbf{n}).
\end{equation}

Note that the boundary condition in Eq.~(\ref{eq:diffusivespinboundary2}) has
the same structure as that for a boundary between an normal metal and an s-wave
superconductor. The reason is that only the s-wave component of the anomalous
Green function survives in the normal metal at distances larger than $l$ from
the boundary. The difference however is that in our case the effective barrier
transparency $t$ in Eq.~(\ref{eq:t2}) depends on the spin-flip tunneling
amplitude $t_s$, and on the vector  $\mathbf{d}$ characterizing the spin
orientation of the triplet order parameter. The phase of the effective s-wave
anomalous Green function  (\ref{eq:eff_greenfunction}), $\chi_0+
\varphi_\mathbf{d}$, also depends on the orientation of the spin vector
$\mathbf{d}$ in the xy-plane.

The aforementioned analogy enables one to treat the proximity effect in  normal
metal- p-wave superconductor systems in the diffusive regime as proximity effect
in an effective s-wave superconductor problem, in which the phase of the s-wave
order parameter and the barrier transparency depend on the spin orientation of
the p-wave condensate.

It is convenient to recast the boundary condition
Eq.~(\ref{eq:diffusivespinboundary2}) in terms of the parametrization in
Eq.~(\ref{eq:G^R_theta}). In our setup, see Fig.~\ref{fig:geometry},  the phase
$\chi(\mathbf{r})$ of the anomalous Usadel Green function (\ref{eq:G^R_theta})
is uniform in space and equal to the phase of the effective s-wave
order parameter, $\chi(\mathbf{r})=\varphi_\mathbf{n} + \chi_0$. The boundary condition for the angle $\theta(\mathbf{r}^N)$ becomes
\begin{equation}\label{eq:bc-theta}
	D\partial_z\theta(\mathbf{r})\big|_{\mathbf{r}=\mathbf{r}^N}
	=2t\cos
	\left[
	 \theta(\mathbf{r}^N)
	\right].
\end{equation}
The Keldysh component of the boundary condition in
Eq.~(\ref{eq:diffusivespinboundary2}) gives the following
boundary condition for the even part of the distribution function:
\begin{eqnarray}
	D\cosh^2\theta_I(\mathbf{r})\partial_z f_1(\mathbf{r})
	|_{\mathbf{r}=\mathbf{r}^N}
	&=&2t\Gamma_\epsilon  f_1(\mathbf{r}^N)\label{eq:bc-distribution}.
\end{eqnarray}
Here we assumed that $f_1(\mathbf{r}^S)=0$ is zero inside superconductor and
introduced the notation
\begin{equation}\label{eq:Gamma_epsilon_def}
    \Gamma_\epsilon =\cosh\theta_I(\mathbf{r}^N) \sin\theta_R(\mathbf{r}^N).
\end{equation}

The set of equations (\ref{eq:sinegordon}) and
(\ref{eq:even-distribution-theta}) along with the boundary conditions
(\ref{eq:bc-theta}) and (\ref{eq:bc-distribution}) gives a description of
electron transport in diffusive metal--p-wave superconductor systems.  Below we
apply these equations to our device geometry.

\section{Resistance of p-wave superconductor- normal metal junction}
\label{sec:solutions}

We consider the geometry in  which the superconductor fills the $z<0$ half space
and the normal metal occupies the $z>0$ half space, see Fig.~\ref{fig:geometry}.
At weak tunneling the Green function in the superconductor is practically
unaffected by the presence of the tunneling barrier. On the other hand, the low
energy properties of the normal metal are significantly affected by the
proximity effect. The singlet Usadel Green function (\ref{eq:G_singlet}) in the
normal metal is described by the set of equations (\ref{eq:G^R_theta}),
(\ref{eq:sinegordon}), (\ref{eq:even-distribution-theta}) with the boundary
conditions (\ref{eq:bc-theta}) and (\ref{eq:bc-distribution}).

The solution of Eq.~(\ref{eq:sinegordon})  satisfying   the condition $\lim_{z\to \infty} \theta(z) = 0$,   has the form
\begin{equation}
	\theta(\epsilon,z)
	=4\arctan
	\left[
		\exp
		\left(
			\beta_\epsilon+(i-1)\frac{z}{L_\epsilon}
		\right)
	\right].
	\label{eq:theta_solution}
\end{equation}
Here $\beta_\epsilon$ is an energy-dependent integration constant. Its value is
determined from the boundary condition in Eq.~(\ref{eq:bc-theta}), which gives
\begin{equation}
	\cosh\beta_\epsilon-\frac{2}{\cosh\beta_\epsilon}
	=(1-i)\frac{L_{t}}{L_\epsilon}
	\label{eq:BC_cosh}
\end{equation}
where
\begin{equation}
	L_{t}=\frac{D}{t},\quad
	L_\epsilon=\sqrt\frac{D}{\epsilon}.
\end{equation}
The algebraic equation (\ref{eq:BC_cosh}) has multiple solutions for the
integration constant $\beta_\epsilon$. The physical solution must satisfy the condition
 $\lim_{\epsilon \to 0}\theta(\epsilon ,z=0+)=\pi/2$, which gives
\begin{equation}\label{eq:exp_beta}
	e^{\beta_\epsilon}=\frac{\alpha+\sqrt{\alpha^2+8}}{2}-
	\frac{1}{2}\sqrt{(\alpha+\sqrt{\alpha^2+8})^2-4},
\end{equation}
where we introduced the notation $\alpha=(1-i)L_t/L_\epsilon$.

\begin{figure}[h!]
	\centering
	\includegraphics[width=0.5\textwidth]{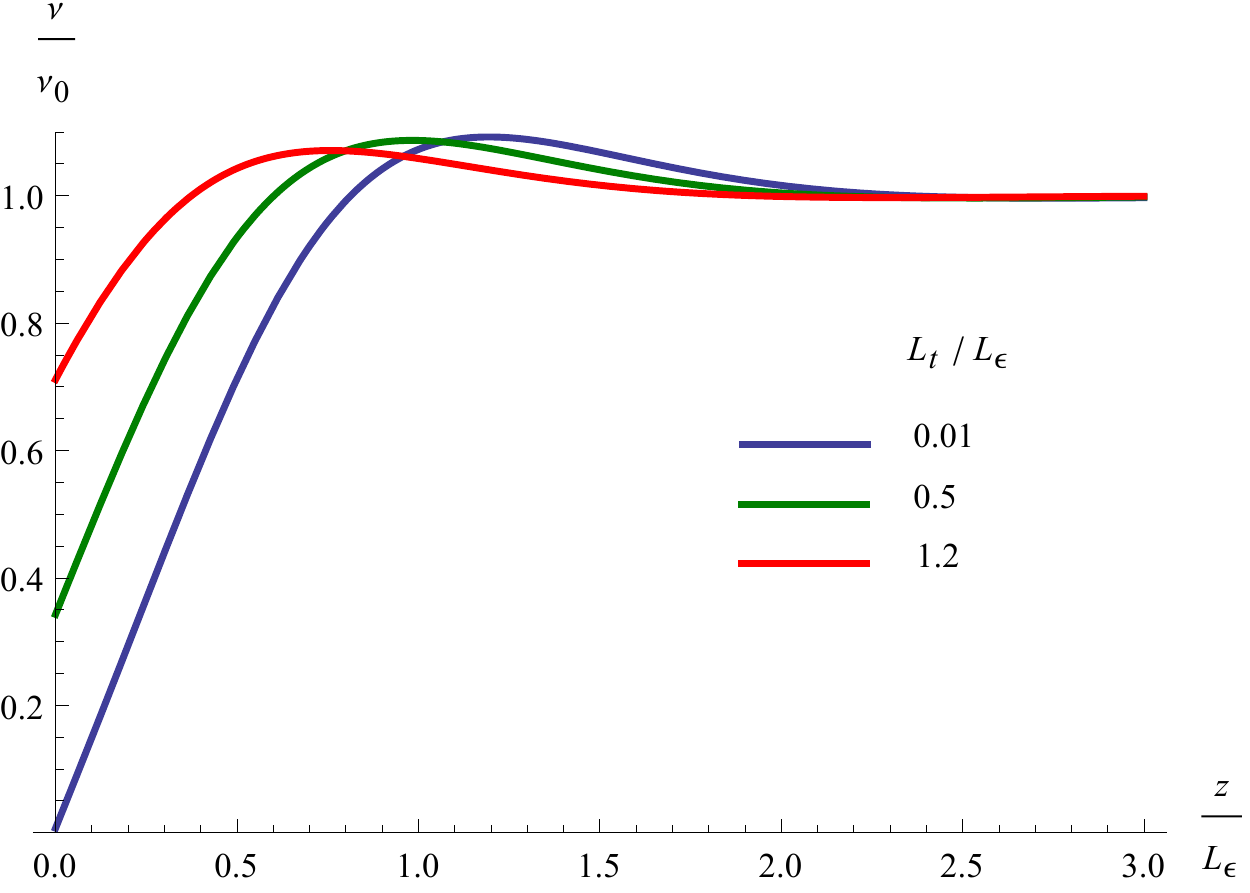}
	\caption{(color online) Density of states in the normal metal as a function of the distance
		from the superconductor-normal metal boundary for different temperatures:
		$L_t/L_\epsilon=0.01$ (blue), $L_t/L_\epsilon=0.5$ (green), and $L_t/L_\epsilon=1.2$ (red).}
	\label{fig:dos_bulk}
\end{figure}

An important aspect  of the solution Eq.~(\ref{eq:theta_solution}) is that in
the normal metal, at small values of $\epsilon$ and at small distances from the
boundary, $\theta(z)\approx \pi/2$ which is the same as in the bulk of the
superconductor.  In particular, it means that at small energies the density of
states in metal is strongly suppressed at distances smaller than $L_{\epsilon}$.
The full spatial dependence of the density of states $\nu(\epsilon,z)$ may be
obtained by substituting the solution (\ref{eq:theta_solution}),
(\ref{eq:exp_beta}) of the Usadel equation into Eqs.  (\ref{eq:dos}), and
(\ref{eq:G^R_theta}).  In Fig.~\ref{fig:dos_bulk} we have plotted  the result as
a function of $z/L_\epsilon$ for different values of $L_{t}/L_\epsilon$.

Note that the effective diffusion constant for the distribution function $f_1$
is determined by the imaginary part of $\theta$, see
Eq.~(\ref{eq:even-distribution-theta}). From the solution
(\ref{eq:theta_solution}) it follows that the imaginary part  $\theta_{I}$ is
close to zero both at $z\gg L_\epsilon$ and $z\ll L_\epsilon$ and has a maximum
at $z\sim L_\varepsilon$ whose value depends on $L_{\epsilon}/L_{t}$.  Therefore
the effective diffusion coefficient in Eq.~(\ref{eq:even-distribution-theta})
approaches  its normal metal value at $z\gg L_\epsilon$ and $z\ll L_\epsilon$.
In the intermediate region $z\sim L_\epsilon$ the diffusion coefficient exceeds
the Drude value.

The differential equation (\ref{eq:even-distribution-theta}) for the
non-equilibrium part of the distribution function has the
first integral, which has the meaning of the conserved partial current density
at a given energy $\epsilon$
\begin{equation}\label{eq:partial_current}
	J_\epsilon\equiv eD\nu_0 \cosh^2\theta_I(z)\partial_zf_1(\epsilon, z).
\end{equation}
The energy dependence of the partial current $J_\epsilon$ can be obtained by
noticing that far away form the boundary the distribution function should have
the form
\begin{equation}\label{eq:f_1_form}
	f_1 (\epsilon, z)= \frac{1}{\cosh^2\frac{\epsilon}{2T}} \frac{eJ_0}{2T \sigma_D} (z -z_0),
\end{equation}
where $\sigma_D=e^2\nu_0 D$ is the Drude conductivity of the normal metal, and
we introduced the current density,
\begin{equation}\label{eq:J_0}
	J_0=\int_{-\infty}^\infty d\epsilon J_\epsilon.
\end{equation}
Substituting Eq.~(\ref{eq:f_1_form}) into Eq.~(\ref{eq:partial_current}) we
obtain the following expression for the partial current
\begin{equation}\label{eq:J_epsilon_result}
	J_\epsilon=
	\frac{J_0}{4T}
	\frac{1}{\cosh^2\frac{\epsilon}{2T}}.
\end{equation}
Using Eqs.~(\ref{eq:partial_current}) and (\ref{eq:J_epsilon_result})  the
solution of Eq.~(\ref{eq:even-distribution-theta}) which satisfies the boundary
condition (\ref{eq:bc-distribution})  and the asymptotic form
(\ref{eq:f_1_form}) at large distances may be written in the form
\begin{equation}
	f_1(\epsilon, z)
	=\frac{eJ_0}{ 2 T \sigma_D \cosh^2\frac{\epsilon}{2T}  }
	\left[
		\frac{L_t}{2\Gamma_\epsilon}+\int_{0}^z\frac{ dz'}{\cosh^2\theta_I(\epsilon, z')}
	\right].
\end{equation}
Here  $\theta_I(z')$ is given by Eqs.~(\ref{eq:theta_solution}) and
(\ref{eq:exp_beta}), and $\Gamma_\epsilon$ was defined in
Eq.~(\ref{eq:Gamma_epsilon_def}).

\begin{figure}
	\includegraphics[width=0.5\textwidth]{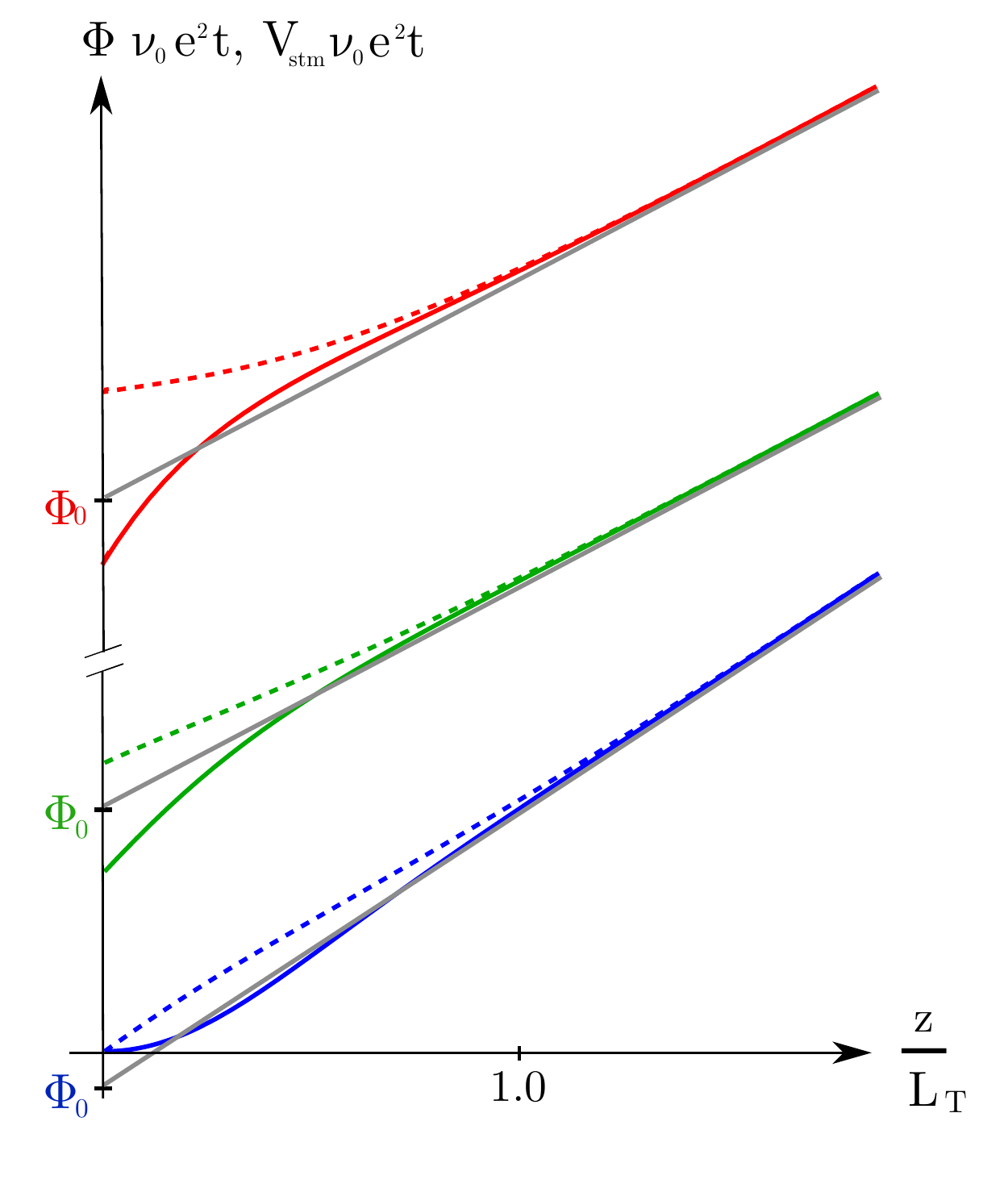}
	\caption{(color online) The spatial variation of the gauge-invariant potential
		$\Phi$ (solid lines) and the compensating voltage $V_{stm}$ at the STM tip
		(dashed lines) on the dimensionless distance $z/L_T$ from the boundary is
		plotted at different temperatures;   $L_t/L_T=0.01$ (blue), $L_t/L_T=1$
		(green) and $L_t/L_T=5$ (red). The solid grey lines represent the large
		distance asymptotes of the gauge invariant potential. Their intercepts with
		the vertical axis for the three values of $L_t/L_T$ are marked by $\Phi_0$
		in the corresponding color. The value of $\Phi_0$ defines the junction
		resistance $R_\infty$ in Eq.~(\ref{eq:R_2_def}). }
	\label{fig:potential_phi}
\end{figure}

Substituting this result in Eq.~(\ref{eq:potential}) we get the following
expression for the gauge invariant potential
\begin{eqnarray}
	\Phi(z)
	&=&\frac{J_0}{\sigma_D}\int_0^\infty \frac{d\epsilon}{2T}
	\frac{\cos\theta_R(\epsilon,z) \cosh\theta_I(\epsilon, z)}{\cosh^2\frac{\epsilon}{2T}}\label{eq:phicapital}
	\nonumber\\
	&&\times
	\left(
		\frac{L_{t}}{2\Gamma_\epsilon}
		+\int_0^z\frac{dz'}{\cosh^2\theta_I(\epsilon, z')}
	\right).
\end{eqnarray}
In Fig. (\ref{fig:potential_phi}), we plotted the dependence of the gauge
invariant potential on the dimensionless distance from the boundary, $z/L_T$,
for different values of the dimensionless barrier transparency parameter,
$L_{t}/L_T$.

One of the important features of transport through the junction is that at low
temperatures  the gauge invariant potential $\Phi(z)$ is significantly
suppressed near the superconductor-normal metal boundary, and is a non-linear
functions of $z$. In particular, the voltage drop across the insulator,
$\Phi(z=0)$, goes to zero in the low temperature limit.

\begin{figure}
	\centering
	\includegraphics[width=0.5\textwidth]{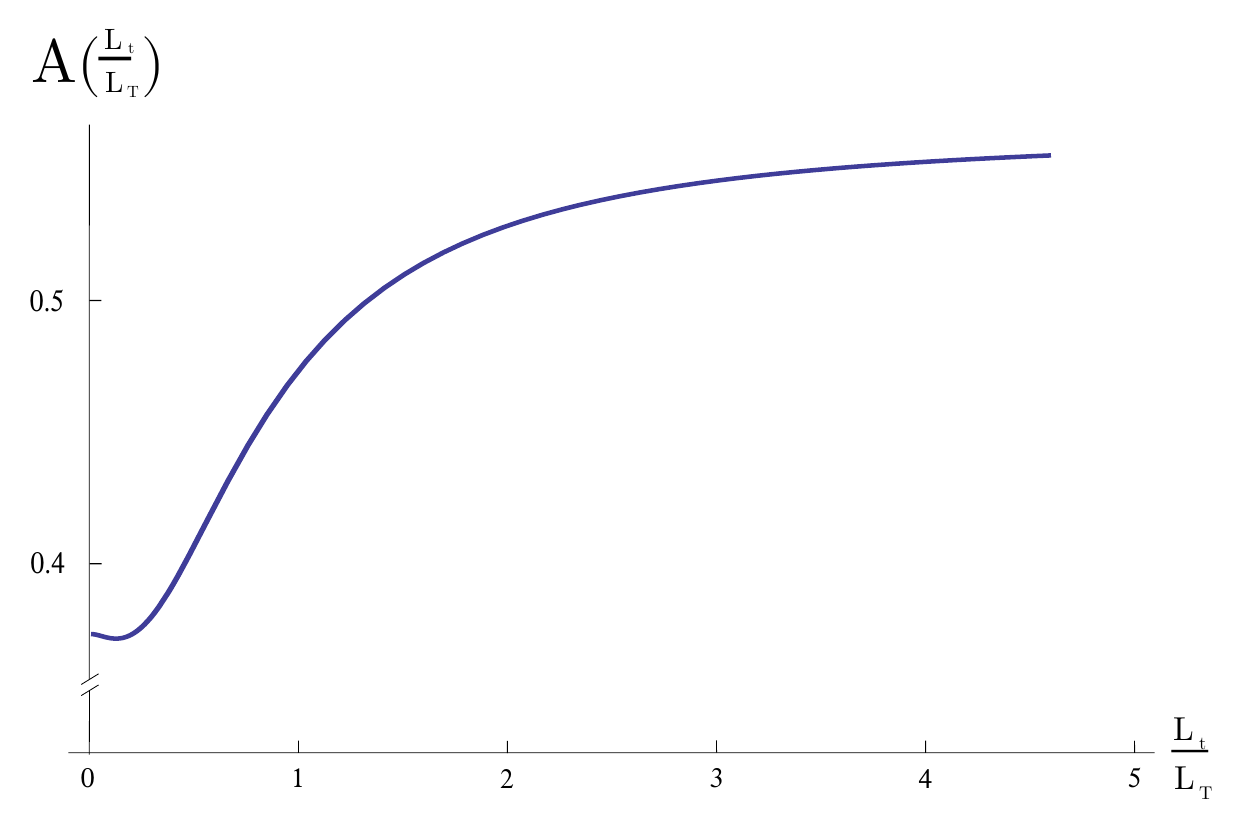}
	\caption{Plot of the function $A(L_t/L_T)$ in Eqs.~(\ref{eq:R_0_result}),
		(\ref{eq:R_0}).}
	\label{fig:R_0}
\end{figure}

Because of the nontrivial spatial distribution of the electric field in the
junction its resistive properties may be characterized in different ways. One
measure of the resistance can be defined in terms of the voltage drop across the
insulating barrier. We define the resistance of the insulating boundary per unit
area as
\begin{equation}\label{eq:R_1_def}
	R_0=\frac{\Phi(z=+0)}{J_{0}}.
\end{equation}
Using Eq.~(\ref{eq:phicapital}) one can expression the boundary resistance $R_0$
per unit area in the form
\begin{equation}\label{eq:R_0_result}
	R_0=\frac{1}{e^{2}\nu_{0}t}\frac{L_{t}}{L_{T}}A\left(\frac{L_{t}}{L_{T}}\right),
\end{equation}
where the dimensionless function $A\left(L_{t}/L_{T}\right)$ is defined by the
following integral
\begin{eqnarray}
	A\left(\frac{L_{t}}{L_{T}}\right)
	&=&\frac{L_T}{L_t}\int_0^\infty \frac{d\epsilon}{4T}
	\frac{\cot\theta_R(\epsilon,0)}{\cosh^2\frac{\epsilon}{2T}}.
	\label{eq:R_0}
\end{eqnarray}
This function is plotted in Fig.~\ref{fig:R_0}. In low and high temperature
limits this expression tends to the following constants; $A(0)\approx 0.37$  and
$A(\infty)\approx 0.53$. As a result in the high and low temperature regimes the
boundary resistance $R_0 \propto \sqrt{T}$.

Note that at low temperatures, $L_T \gg L_{t}$, the magnitude of the jump of
$\Phi(z)$ at the insulator boundary  approaches zero at $T\to 0$. This is very
different from the resistance of the normal metal-insulator-normal metal
junctions where in the presence of a current though the junction
$R_{NIN}=1/e^{2}\nu_{0} \tilde{t}$, where $\tilde{t}\sim t_0^{2}$ is the
transmission coefficient of the insulator.

Another measure of the junction resistance may be obtained by extrapolating the
linear dependence of $\Phi(z)$ at large distances,  $\Phi(z)=  J_0 z/\sigma_D +
\Phi_0$  to the location of the barrier, $z=0$. This is shown by grey solid
lines in Fig.  ~\ref{fig:potential_phi}. The value of the intercept with the
vertical axis, $\Phi_0$, defines the total resistance per unit area of the
junction
\begin{equation}\label{eq:R_2_def}
	R_\infty=
	\frac{\Phi_0}{J_0}.
\end{equation}
Using Eq.~(\ref{eq:phicapital}) we obtain
\begin{equation}
	R_\infty = \frac{1}{e^2t\nu_0}B\left(\frac{L_t}{L_T}\right),
\end{equation}
where the function $B(L_t/L_T)$ is given by the following integral
\begin{equation}
	B= \int_0^\infty\frac{d\epsilon}{2T}\frac{1}{\cosh^2\frac{\epsilon}{2T}}
	\left[
		\frac{1}{2\Gamma_\epsilon}-
		\int_0^\infty \frac{dz'}{L_t} \tanh^2\theta_I(\epsilon,z')
	\right].
	\label{eq:R_inf}
\end{equation}
The first term in the brackets is positive and  represents the contribution of
the insulating boundary.  The second term is negative. It describes the
reduction of the resistance of the normal metal due to the proximity effect.

\begin{figure}
	\includegraphics[width=0.5\textwidth]{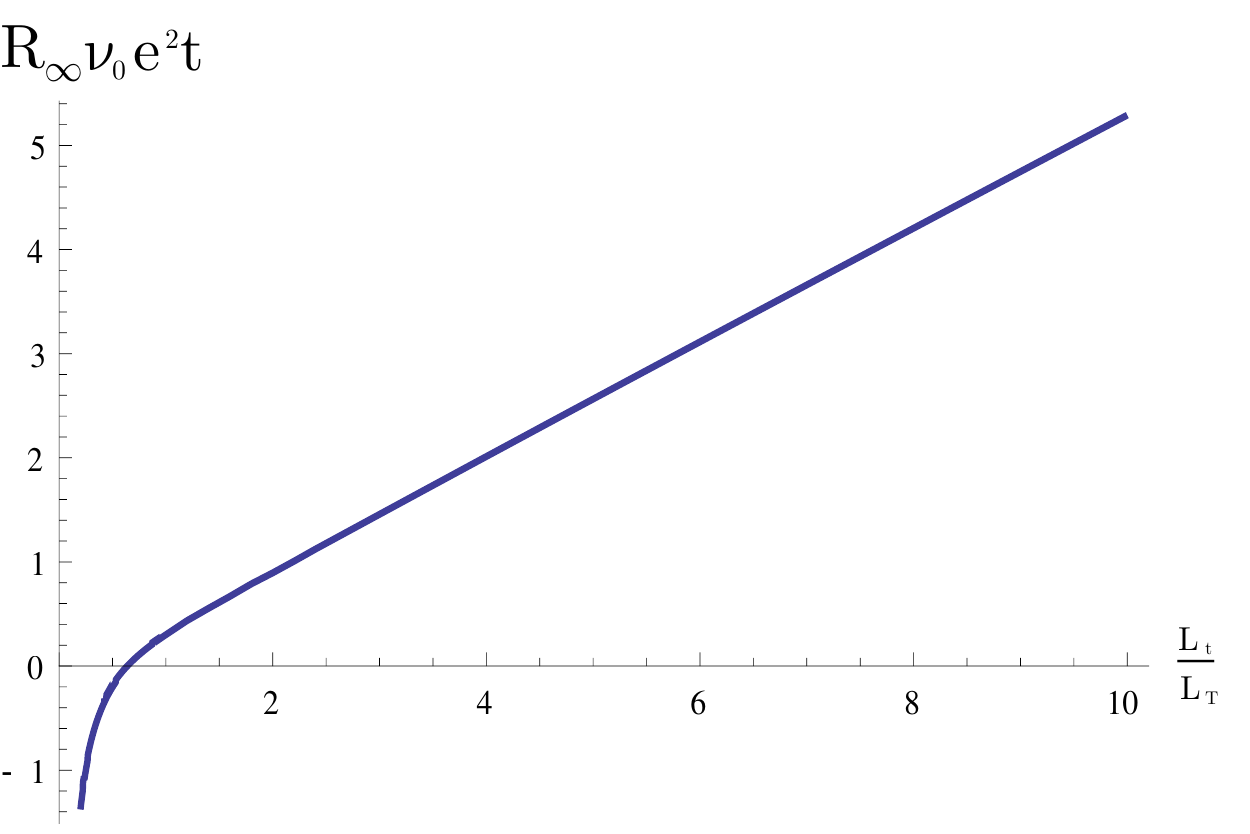}
	\caption{The junction resistance $R_\infty$ per unit area (in units of $1/e^2
		\nu_0 t$) is a plotted as a function of $L_t/L_T$. } \label{fig:R_inf}
\end{figure}

The junction resistance $R_\infty$ is plotted in Fig.~(\ref{fig:R_inf}) as a
function of $L_t/L_T$. At relatively high temperatures $L_t/L_T\gg 1$, junction
resistance $R_\infty$ is dominated by the contribution from the insulating
boundary (first term in Eq.~(\ref{eq:R_inf})). In this case $B(L_t/L_T)\approx
0.53\,  L_t/L_T$,  in agreement with the discussion below Eq.~(\ref{eq:R_0}). In
the low temperature regime, $L_t\ll L_T$, the junction resistance is dominated
by the change in the resistance of the normal metal due to the proximity effect
(second term in Eq.~(\ref{eq:R_inf}) and becomes negative. In this case  the
junction resistance reduces  to
\begin{equation}\label{eq:R_inf_low_T}
	R_\infty =   - \frac{b}{e^2t\nu_0}\frac{L_T}{L_t},
\end{equation}
where the constant $b$ is given by
\begin{eqnarray}\label{eq:gamma_def}
    b&=&
	\int_0^\infty \frac{d\lambda}{2}
	\frac{\lambda^{-1/2}}{\cosh^2\frac{\lambda}{2}}\nonumber	\\
	&&\times\int_0^\infty d\zeta\tanh^2\left[4\, \mathrm{Im} \arctan\left(
			(\sqrt{2}-1)e^{(i-1)\zeta}\right)\right]\nonumber\\
	&\approx&0.39 .
\end{eqnarray}

\subsection{Probing the spatial distribution of the gauge-invariant potential $\Phi (\mathbf{r})$}

Let us now discuss the possibility of experimental observation of the
suppression of $\Phi(x)$ near the junction's boundary by using a scanning tunneling probe.  
We consider the setup
illustrated in Fig.~\ref{fig:geometry}.  

The electron transport between the STM tip and the
metal can be described with the aid of the tunneling Hamiltonian
\begin{equation}
	H_T=\sum_{\mathbf{k}\mathbf{p}}
	\left[
	t_{\mathbf{kp}}
		c_\mathbf{k}^\dagger c_\mathbf{p}+t_{\mathbf{kp}}^*
		c_\mathbf{p}^\dagger c_\mathbf{k}\right].
\end{equation}
Here $c^\dagger$ is an electron creation operator, and $\mathbf{k}$ labels the
states in the STM tip and $\mathbf{p}$ labels the states in the wire.
In the tunneling approximation the STM current can be written in the form
%
%
\begin{eqnarray}\label{eq:voltage}
		I_{stm}(z) &=&
	\frac{g_n}{2e}	\int_{-\infty}^{\infty} d\epsilon
	\cos\theta_R(\epsilon,z) \cosh\theta_I(\epsilon, z)\nonumber\\
	&&\times\left[
		f_1^{stm}(\epsilon)-
		f_1(\epsilon, z)
	\right],
\end{eqnarray}
where $g_n$ is the conductance of the tunneling contact in the normal state. 
The nonequilibrium distribution function in the STM is given by
$f_1^{stm}(\epsilon)=eV_{stm}/2T\cosh^2(\epsilon/2T)$, where $V_{stm}$ is the
STM voltage measured relative to that in the superconductor.

Using Eq.~(\ref{eq:potential}) we can rewrite Eq.~(\ref{eq:voltage}) in the form
\begin{equation}\label{eq:tunneling_current_short}
    I_{stm}(z)=g_n\Phi (z) - g_t(T, z)V_{stm},
\end{equation}
where
\begin{equation}\label{eq:tunneling conductance}
    g_t(T, z)=g_n\int_{0}^{\infty} \frac{d\epsilon}{2T}\frac{\cos\theta_R(\epsilon,z) \cosh\theta_I(\epsilon, z)}{ \cosh^2\frac{\epsilon}{2T}}
\end{equation}
is the   conductance of the tunneling contact.

In the case where the voltage $V_{stm}$ at
the tip  vanishes the value of
the tunneling current through the STM contact is proportional to $\Phi(z)$,
\begin{equation}\label{Istm}
I_{stm}(z)=g_n\Phi(z).
\end{equation}
In particular, $I_{stm}(z)$ is significantly suppressed near the
superconductor-normal metal boundary, reflecting corresponding suppression of $\Phi(x)$.

On the other hand, if $I_{stm}=0$,  we get
\begin{equation}\label{Vstm}
	V_{stm}(z)=\frac{g_n}{g_t(T, z)}\,  \Phi(z),
\end{equation}
where $\Phi(z)$ is given by Eq.~(\ref{eq:phicapital}). The graph of $V_{stm}(z)$
is plotted in Fig.~\ref{fig:potential_phi} by dashed lines for several
temperatures.  It is interesting to note that, in contrast to the gauge
invariant potential, Eq.~(\ref{Istm}), the compensating STM voltage in
Eq.~(\ref{Vstm})  does not exhibit the aforementioned suppression near the
boundary at low temperatures, $L_T \gg L_t$.  The slope $d V_{stm}(z)/dz $
remains practically the same as in the normal metal in the absence of
superconductor, both at  $z\ll L_{T}$ and at  $z\gg L_{T}$.   The reason is that
the conductance of the tunneling barrier between the STM and the metal, $g_t(T, z)$,   reflects the suppression of the single particle density of
states in the metal, as described by Eq.~(\ref{eq:tunneling conductance}). This nearly  cancels the suppression
of $\Phi(z)$ in Eq.~(\ref{Vstm}).

\section{Conclusions}\label{sec:conclusion}

We show that the low temperature resistance of the p-wave
superconductor-diffusive normal metal junctions is controlled by the spin-orbit
interaction. As a result the junction resistance, tunneling density of states in
the metal  and other transport properties of the device exhibit a strong
dependence on the angle between the vector ${\bf d}$ characterizing the spin
part of the superconducting wave function, and the normal to the surface of the
junction.  In particular, the s-wave component of the proximity effect in metal
vanishes when ${\bf d}$ is parallel to the c-axis.

The absence of the corresponding  dependence of the Knight shift on the angle
between ${\bf d}$ and the c-axis in $Sr_{2}RuO_{4}$ crystals is one of the
problems in the interpretation of $Sr_{2}RuO_{4}$ as a conventional p-wave
superconductor.  This fact was attributed to weakness of the  spin-orbit
interaction in $Sr_{2}RuO_{4}$.\cite{knight} We would like to point out that the
resistance of the junction should be strongly dependent on the angle between
${\bf d}$ and ${\bf z}$ even in the case of weak spin-orbit interaction.
Therefore the measurement of this effect could clarify the situation.

Another consequence of the sensitivity of the proximity effect to the
orientation of the condensate spin is that a current passing across such a
junction leads to spin accumulation inside the p-wave superconductor (although
inside the proximity region no spin accumulation occurs).

We also would like to mention that the boundary conditions
Eq.~(\ref{eq:diffusivespinboundary2}) can be used to describe the Josephson
effect in junctions consisting of two p-wave superconductors separated by a
diffusive normal metal.  The structure of boundary conditions
(\ref{eq:diffusivespinboundary2}) is similar to those of  for s-wave
superconductor-normal metal junction. Therefore the supercurrent for the p-wave
case may be obtained from the conventional formulas for the s-wave case if we
substitute the phase difference in the s-wave case with
$\phi_\mathbf{d}+\chi_0$, see Eqs.~(\ref{eq:eff_greenfunction}) and
(\ref{eq:d_angles}), and the transmission coefficient with $t$.

An important consequence of the proximity effect near the superconductor-normal
metal boundary is the suppression of the Hall effect in the metal near the
superconducting boundary.  Qualitatively, this suppression is related to the
fact that, due to proximity effect, at low energies the quasiparticle wave
functions in metal  are a coherent superposition of electron and hole wave
functions, and the effective charge of the quasiparticles approaches zero at
$\epsilon\rightarrow 0$.  The presented above scheme of calculation of the
electronic transport was derived in zeroth order in $\omega_{c}\tau$, where
$\omega_c$ is the cyclotron frequency and $\tau$ is the elastic mean free time.
In this approximation the electron wave functions near the Fermi surface are
electron-hole symmetric, which yields a vanishing Hall effect. To describe Hall
effect one has to add to the expression for the current a term linear in
$\omega_{c}\tau$,\cite{ZhouSpivak}
\begin{equation}
		\mathbf{J}_H\propto \omega_{c}\tau {\bf b}\times\int d\epsilon
		\cos\theta_{R}\cosh^3\theta_{I}
		\boldsymbol{\nabla} f_{1}
\end{equation}
here $\omega_{c}$ is the cyclotron frequency, and  ${\bf b}$ is the unit vector
in the direction of the magnetic field. Since the magnitude of the proximity
effect is controlled by $t$, which is proportional to
$\sin\vartheta_\mathbf{d}$, the Hall conductance is expected to have a strong
dependence on the orientation of the order parameter, $\mathbf{d}$. Since the
latter may be oriented by the external magnetic field, both the
magnetoresistance of the junction and the Hall resistance are expected to be
strongly anisotropic with respect to orientation of the magnetic field.

Finally, we note that our results hold for more general realizations of
$p_x+ip_y$ order parameter in superconductors with complicated topology of the
Fermi surface, such as the one proposed in  Ref.~\onlinecite{kivelson}.

This work was supported by the US Department of Energy through the grant  DE-FG02-07ER46452. B. S. thanks the International Institute of
Physics (Natal, Brazil) for hospitality during the completion of the paper.

\end{document}